\documentclass[aps,twocolumn,showpacs,superscriptaddress]{revtex4}

\usepackage{graphicx}
\usepackage{subfigure}
\usepackage[latin1]{inputenc}
\usepackage{amsmath}
\usepackage{amssymb}
\usepackage{color}

\newcommand{\nc}{\newcommand}
\nc{\bra}[1]{\langle #1|}
\nc{\ket}[1]{|#1\rangle}
\nc{\braket}[1]{\left\langle #1 \right\rangle}
\nc{\equ}[1]{\begin{eqnarray*}#1\end{eqnarray*}}
\nc{\equn}[1]{\begin{eqnarray}#1\end{eqnarray}}
\nc{\dagg}{^{\dagger}}
\nc{\conj}{^{*}}
\nc{\dx}[1]{\, \mathrm{d} {#1}}
\nc{\Dx}[1]{\mathcal{D} {#1} \,}
\nc{\la}{\langle}
\nc{\ra}{\rangle}
\nc{\Tr}{\text{Tr} \,}
\nc{\e}{\text{e}}
\nc{\Id}{\mathbb{1}}
\nc{\eps}{\varepsilon}
\nc{\der}[2]{\delta \frac{{#1}}{\delta {#2}}}
\nc{\pder}[2]{\frac{\partial {#1}}{\partial {#2}}}
\nc{\bigO}{\mathcal{O}}

\nc{\eq}[1]{Eq.(\ref{#1})}
\nc{\chap}[1]{Chapter \ref{#1}}
\nc{\sect}[1]{Section \ref{#1}}
\nc{\fig}[1]{Fig.\ref{#1}}
\nc{\Fig}[1]{Figure \ref{#1}}
\nc{\tabl}[1]{Table \ref{#1}}
\nc{\app}[1]{Appendix \ref{#1}}

\nc{\eg}{\emph{e.g.} }
\nc{\ie}{\emph{i.e.} }

\nc{\alg}[1]{\textcolor{blue}{#1}}

\begin{document}


\title{Anomalous Paths in Quantum Mechanical Path-Integrals}


\author{Arne L. Grimsmo}\email{arne.grimsmo@ntnu.no}
\affiliation{Department of Physics, 
             The Norwegian University of Science and Technology, N-7491 Trondheim, Norway}
\affiliation{Department of Physics, 
            The University of Auckland, Private Bag 92019, Auckland, New Zealand}
\author{John R.~Klauder}\email{klauder@phys.ufl.edu}
\affiliation{
Departments of Physics and Mathematics,
University of Florida,
Gainesville, FL 32611, U.S.A.}
\author{Bo-Sture K. Skagerstam}\email{bo-sture.skagerstam@ntnu.no}
\affiliation{Department of Physics, 
             The Norwegian University of Science and Technology, N-7491 Trondheim, Norway}
\affiliation{CREOL, The College of Optics and Photonics at the University of Central Florida, 4000 Central Florida Boulevard, Orlando, Florida 32816, USA}


\begin{abstract}
We investigate modifications of the discrete-time lattice action, for a quantum mechanical particle in one spatial dimension, that vanish in the naïve continuum limit but which, nevertheless,  induce non-trivial effects due to quantum fluctuations. These effects are seen to modify the geometry of the paths contributing to the path-integral describing the time evolution of the particle, which we investigate through numerical simulations. In particular, we demonstrate the existence of a modified lattice action resulting in paths with any fractal dimension, $d_f$, between one and two. We argue that $d_f=2$ is a critical value, and  we exhibit  a type of lattice modification where the fluctuations in the position of the particle becomes independent of the time step, in which case the paths are interpreted as superdiffusive L\'{e}vy flights.  We also consider the jaggedness of the paths, and show that this gives an independent classification of lattice theories.
\end{abstract}

\maketitle
\section{Introduction}
The path-integral representation of the amplitude $\braket{x',t'|x,t}$ for a quantum mechanical particle of mass $m$ moving in a local potential $V(x)$ is usually written as a limit of a multi-dimensional integral \cite{FeynmanHibbs}:
\begin{align} \label{eq:intro:Z}
Z \equiv \braket{x',t'|x,t} = \lim_{N\to\infty} \mathcal{N} \int \dx{x}_1 \dots \dx{x}_{N-1} \e^{-S_N},
\end{align}
where we have changed to imaginary time ($t \to -it$) and set $\hbar = 1$. Here $\mathcal{N} = (m/2\pi a)^{N/2}$ and $S_N$ is the discrete-time action which should approach the classical continuum action $S$ as the lattice constant $a \equiv (t_f-t_i)/N$ goes to zero, \emph{i.e.},
\begin{align}
\lim_{N\to \infty} S_N = S = \int_{t_i}^{t_f} \dx{t} \left[\frac{1}{2} \dot{x}^2 + V(x)\right].
\end{align}
We have chosen units such that the mass $m$ of the particle is one.  The particular choice
\begin{align} \label{eq:intro:naive}
S_N \equiv \sum_{k=0}^{N-1}S_k = \sum_{k=0}^{N-1}a\left[\frac{1}{2}\left(\frac{\Delta x_k}{a}\right)^2 + V(x_k)\right],
\end{align}
where $\Delta x_k \equiv x_{k+1}-x_k$, with a time-step $dt \to \Delta t \equiv a$, is referred to as the naïve discretization of the classical action $S$, and has, for example, been used in modeling time as a discrete and dynamical variable \cite{Lee_1983}.  The choice Eq.(\ref{eq:intro:naive}) is, however,  by no means unique and the ambiguity of the discretization has been investigated previously by, \emph{e.g.}, Klauder et al. in Ref.\cite{Klauder}. As an interesting example, it has also  been shown that adding terms proportional to $a \Delta x_k^{2n}$, as $a\to0$ ($n=1,2,\dots$), to each term $S_k$  in the sum in Eq.(\ref{eq:intro:naive}) permits a radical speedup of the convergence in Monte Carlo-simulations \cite{Bogojevic}. Classically, one expects $\Delta x_k/a \to \dot{x}$ to be well-defined as $a \to 0$ and thus $S_k = \mathcal{O}(a)$, and, as was noted in Ref.\cite{Bogojevic}, one would have $a \Delta x_k^{2n} \to a^{2n+1} \dot{x}^{2n} = \mathcal{O}(a^{2n+1})$, which clearly vanish in the $a\to0$ limit. We will refer to these considerations as the ``naïve continuum limit'' in the following.

As was pointed out in Ref.\cite{Klauder} the argumentation above is, however, not true  for quantum mechanical paths, as one expects $\Delta x_k = \bigO(\sqrt{a})$ in accordance with the Itô calculus for a Wiener process, and thus the action then contains terms $S_k$ of order one. Modifications as those considered in Ref.\cite{Bogojevic} still vanish, but only as fast as $\bigO(a^{n+1})$. This implies no difficulty for the numerical speedup procedure, but in general, it is clear that one must take care when modifying the action in the presence of quantum fluctuations.

We now wish to expand on the work from Ref.\cite{Klauder} and proceed to study precisely those modifications to the discrete action that vanish in the naïve limit, but might induce non-trivial effects when quantum fluctuations are taken into account. We will show that not only can non-vanishing local potentials be induced by such alterations, as was shown in Ref.\cite{Klauder}, but the situation is further complicated in that the size of quantum fluctuations can be changed under the modified lattice theory, such that no naïve assumptions on the continuum limit can be made. 
This can be seen to manifest itself in the geometrical properties of the paths contributing to the path-integral, and as we will see shortly, can generate both sub- and super-diffusive behaviour.

\section{Geometry of path-integral trajectories}

We now quickly review two useful measures that will be used to quantify the geometry of relevant paths in the path-integral. The geometry of path-integral trajectories has been investigated previously, in particular by Kröger et al. in Ref.\cite{Kroger}, where a fractal dimension was defined and found both analytically and numerically for local and velocity-dependent potentials. More recently a complementary property termed ``jaggedness'' was identified by Bogojevic et al. in Ref.\cite{Bogojevic2}. Both of these measures signify the relevance of different paths as to what degree they contribute to the total path-integral.

To define the fractal dimension, $d_f$, for path-integral trajectories, we recall that the fractal dimension for a classical path can be defined in the following way: We first define a length of the path, $L(\epsilon)$, as obtained with some fundamental resolution $\epsilon$. This can, for example,  be done by making use of a minimal covering of the path with ``balls'' of diameter $\epsilon$ such that $L(\epsilon) = N(\epsilon) \times \epsilon$,  where $N(\epsilon)$ is the number of balls. A fractal dimension can then be defined as the unique number $d_f$ such that $L(\epsilon) \sim \epsilon^{1-d_f}$ as $\epsilon \to 0$ \cite{mandelbrot_82}. For path-integral trajectories, a total length can be defined as $\braket{L} = \braket{\sum_k |\Delta x_k|}$, and the role of $\epsilon$ will be played by the expected absolute change in position, $\braket{|\Delta x_k|}$,  over one small time step $\Delta t \simeq a$. Here $\braket{\cdot}$ denotes the quantum-mechanical average  using the probability distribution obtained from Eq.(\ref{eq:intro:Z}).  For a typical value, $|\Delta x|$, of $\braket{|\Delta x_k|} $, say $|\Delta x| \simeq (\Delta t)^{1/ \gamma}$, we then have that $\braket{L} \simeq N |\Delta x| \simeq T |\Delta x|^{1-\gamma}$ since $N \simeq T/\Delta t $, with $T=t_f -t_i$.  We then conclude that $d_f=\gamma$. 
The fractal dimension can therefore be obtained through a scaling with the number of lattice sites $N$, as $N \to \infty$, with $T= N\Delta t\simeq Na$ held fixed, \emph{i.e.},
\begin{align}\label{eq:intro:df_def}
&\braket{L} \sim N^{1-1/d_f}  ,
\end{align}
for sufficiently large $N$.
This is also the definition made use of in Ref.\cite{Kroger}, and is a measure of how the increments $\braket{|\Delta x_k|}$ scale with the time step $\Delta t \simeq a$. In the spirit of anomalous-diffusion considerations (see e.g. Refs.\cite{Metzler_2000}), we will refer to those paths with a fractal dimension $d_f < 2$, as defined above, as sub-diffusive, reflecting that they spread in space at a slower than normal rate. Similarly those paths with $d_f > 2$ are referred to as super-diffusive, which  then corresponds to L\'{e}vy flights (see e.g. Refs.\cite{Frisch_1995}). 

A remark on the physical interpretation of $d_f$ is in order before we proceed. The length $\braket{L}$  defined above is not necessarily an experimentally observable length. It gives us, however, an insight into the nature of how the geometry of those paths with a non-zero measure  change under modification of the lattice action.  The definition of a fractal dimension for the \emph{physical} path of a quantum mechanical particle must necessarily involve considerations  of a measuring apparatus, as was done by Abbott and Wise \cite{AbbottWise}. Inclusion of quantum measurements in a path-integral framework has been discussed in the literature (see e.g. Ref.\cite{Mensky}), but will not be considered in this work. 

It is well known that the paths contributing to the path-integral, \eq{eq:intro:Z}, are continuous but non-differentiable. Indeed, using a partial integration, Feynman and Hibbs \cite{FeynmanHibbs} showed that for any observable $F$ the identity
\begin{align}
\left\langle \frac{\delta F}{\delta  x_k} \right\rangle = \left\langle F \frac{\delta S}{\delta  x_k} \right\rangle
\end{align}
holds. In the case $F = x_k$ this leads to
\begin{align} \label{eq:intro:dx2}
\braket{\Delta x_k^2} = {\cal O}(a) \, ,
\end{align}
for the lattice action Eq.(\ref{eq:intro:naive}) and for sufficiently small $a$, and where we from now on assume that expressions like $\braket{x_kdV(x_k)/dx_k}$ are finite.
Hence, we expect $\braket{|\Delta x_k|} \propto 1/\sqrt{N}$ and $\braket{L} \propto \sqrt{N}$ corresponding to a fractal dimension of $d_f = 2$, which has been confirmed numerically in Ref.\cite{Kroger}.

The second measure we will use to describe the relevant paths in the path-integral, is the ``jaggedness'', $J$, defined in Ref.\cite{Bogojevic2}, which counts the number of maxima and minima of a given path:
\begin{align}
J = \frac{1}{N-1} \sum_{k=0}^{N-2} \frac{1}{2} [1-\left \langle \text{sgn}(\Delta x_k \Delta x_{k+1}) \rangle \right],
\end{align}
with $J \in [0,1]$. It is a  measure of the correlation between $\Delta x_k$ and $\Delta x_{k+1}$ with $J = 1/2 + \bigO(a)$ for completely uncorrelated increments. We therefore expect the jaggedness to be invariant under modifications only altering nearest neighbor interactions on the lattice. Below we will consider the average value of $J$ for sub and super-diffusive paths.

\section{Sub-diffusive paths}
\begin{figure}
\includegraphics[width=\columnwidth]{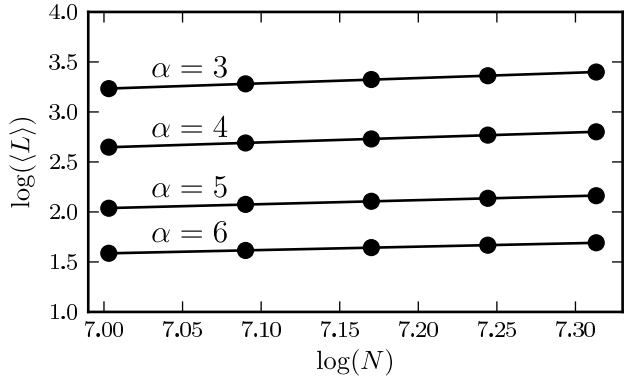}
\caption{\label{fig:intro:LvN_sub}Scaling of the average length $\braket{L}$ as a function of lattice sites, $N$, for the lattice action given in \eq{eq:intro:Sv}, with $\xi=2$ and various $\alpha$. $\log(\braket{L})$ was fitted to $\beta \log(N) + b$. The values for $\alpha$ are $3, 4, 5$ and $6$, starting from the top line and descending. Statistical error bars are not visible in the figure.}
\end{figure}

\begin{figure}
\includegraphics[width=\columnwidth]{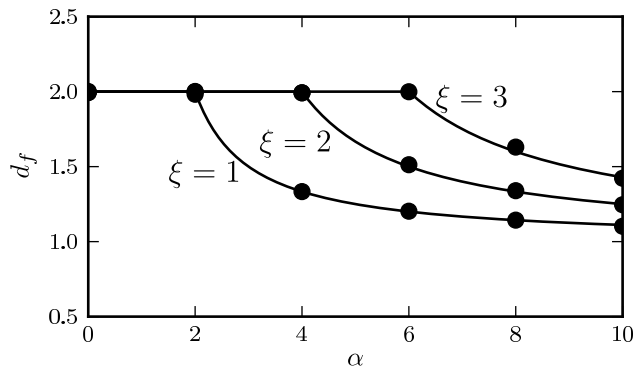}
\caption{\label{fig:intro:df_sub}The fractal dimension, $d_f$, as a function of $\alpha$, for $\xi=$ 1, 2 and 3, for the action defined in \eq{eq:intro:Sv}. The dots are numerical results, and the solid lines represent the expected theoretical values according to $d_f = \alpha/(\alpha - \xi)$, with $\xi=3$ (the top line), $\xi=2$ (the middle curve), and $\xi=1$ (the lowest curve).} 
\end{figure}

Sub-diffusive paths, as defined here, were discovered to be the contributing paths in the presence of a velocity dependent potential, $V_0 |v|^\alpha$, in Ref.\cite{Kroger}. We will here consider a similar modification, that in fact \emph{vanish} in the naïve continuum limit, yet changes the geometry of the paths when quantum fluctuations are taken into account:
\begin{align}\label{eq:intro:Sv}
S_k \to S_k + g a^{\xi}\, \left|\frac{\Delta x_k}{a}\right|^\alpha,
\end{align}
where $g$ is a coupling constant, $\xi \ge 1$ and $\alpha \ge 0$. The last term is identical to the modification considered in Ref.\cite{Kroger} for $\xi=1$, but naïvely vanishes for any $\xi > 1$. Due to quantum fluctuations, however, \eq{eq:intro:dx2} must be replaced by
\begin{align}
\frac{1}{a} \braket{\Delta x_k^2} + g \alpha a^{\xi-\alpha}\braket{|\Delta x_k|^\alpha} = {\cal O}(1) ,
\end{align}
showing that for $\alpha > 2 \xi$ the last term dominates, and we expect $\braket{|\Delta x_k|} \propto a^{(\alpha-\xi)/\alpha}$, corresponding to a fractal dimension of $d_f = \alpha/(\alpha-\xi)$. For $\alpha \le 2\xi$ we still have $d_f = 2$ showing that $2\xi$ is a critical point for the fractal dimension as a function of $\alpha$. For $\xi = 1$ this reproduces the results from \cite{Kroger}. In \fig{fig:intro:LvN_sub} we show how the length $\braket{L}$ scales with the number of lattice sites $N$ for various $\alpha$ and $\xi=2$. The results are produced numerically by standard Monte Carlo methods \cite{Kroger,Hastings,C&F}. From this scaling one can find the fractal dimension according to \eq{eq:intro:df_def}. In \fig{fig:intro:df_sub} we have extracted the fractal dimension as a function of $\alpha$ numerically for $\xi =$ 1, 2 and 3. We see that the numerical results fit well to the expected values of $d_f = 2$ for $\alpha \le 2\xi$ and $d_f = \alpha/(\alpha-\xi)$ for $\alpha > 2\xi$, shown as solid lines in the figure.


\section{Super-diffusive paths}
Consider now modifications of the form
\begin{align} \label{eq:intro:fS}
S_k \to f(S_k),
\end{align}
for some analytical function, $f(x)$, with the constraint $f(x) = x$ as $x \to 0$, in order to reproduce the classical limit. This constitutes a large class of local modifications---\emph{i.e}, only influencing nearest-neighbor couplings on the lattice---that have the same naïve $a \to 0$ limit. 

We will also assume, if required,  that there exists some large distance infrared cutoff so that the integral
\begin{align}
\psi(x_{k+1}) = N \int \dx{x_k} \e^{-f(S_{k})} \psi(x_k)\, ,
\end{align}
exists, describing the evolution of the wave-function $\psi(x)$ over a small time step, $a$, under the modified lattice action. For reasons of simplicity, we assume the infrared regularization $\psi(x_k) = 0$ for $|x_{i+k}-x_k| \ge L$, for some sufficiently large $L$.

Through a straightforward renormalization procedure (see the Appendix) we are able to write down an effective action that is equivalent to the modification, \eq{eq:intro:fS}, in the continuum limit. Remarkably, the so obtained effective action can formally be written in the naïve form of \eq{eq:intro:naive}, i.e.
\begin{align}\label{eq:intro:naive_eff}
S_N \rightarrow S_N^\text{eff} = a \sum_{k=0}^{N-1} \left[ \frac{1}{2 s[f]} \left(\frac{\Delta x_k}{a}\right)^2 + g[f]V(x_k)\right],
\end{align}
where
\begin{align}\label{eq:intro:sfgf1}
s[f] &= \frac{ \int_{\Omega} \dx{y} y^2 \exp\left(-f\left(\frac{y^2}{2}\right)\right) }{\int_{\Omega} \dx{y} \exp\left(-f\left(\frac{y^2}{2}\right)\right)}\,\, , 
\end{align}
and
\begin{align}\label{eq:intro:sfgf2}
g[f] &= \frac{ \int_{\Omega} \dx{y} f'\left(\frac{y^2}{2}\right) \exp\left(-f\left(\frac{y^2}{2}\right)\right) }{\int_{\Omega} \dx{y} \exp\left(-f\left(\frac{y^2}{2}\right)\right)}\,\, .
\end{align}
Here the integrals run over the domain $\Omega$, given by $-L/\sqrt{a} < y < L/\sqrt{a}$. Hence the integrals become independent of the infrared cut-off, $L$, in the $a \to 0$ limit. As discussed in the Appendix, this can be interpreted as a renormalization of the particle's mass and potential, and will in general be finite or infinite in the limit $a \to 0$, depending on the form of $f$. 
With the modification \eq{eq:intro:fS}, \eq{eq:intro:dx2} must, however,  be replaced by
\begin{align} \label{eq:intro:dx2_mod}
\braket{f'(S_k)\Delta x_k^2} = {\cal O}(a),
\end{align}
potentially changing how $\braket{|\Delta x_k|}$ scales with $a$ and thus the fractal dimension $d_f$ as defined above. Similarly, for the equivalent effective counterpart \eq{eq:intro:naive_eff}, we see that the scaling can be written
\begin{align} \label{eq:intro:dx2_sf}
\braket{\Delta x_k^2} = s[f]a \, ,
\end{align}
and therefore all modifications to the short-time scaling are contained in the functional $s[f]$.
For a function $f(x)$ that is bounded, however, $s[f]$ diverges like $L^2/a$ as $a$ goes to zero, and therefore $\braket{\Delta x_k^2} \simeq L^2$ in terms of the infrared cutoff $L$. In this case we therefore expect that the particle can make arbitrarily large jumps, independent of $a$. This behavior can be interpreted, at least formally, as an infinite fractal dimension for the particle's path since $\braket{\Delta x_k^2} \simeq a^{2/d_f}$, and is typical for \emph{any} such $f$.   Such paths are analogous to Poisson paths, such as appear in Ref.\cite{Klauder_93},  which involve paths with continuous segments joined by jumps whose magnitude is drawn from a well defined distribution at time intervals, again, with a suitable distribution.

We illustrate these features in terms of the following family of lattice modifications, defined through \eq{eq:intro:fS},
\begin{align} \label{eq:intro:falpha}
f\equiv f_\gamma(x) =
\begin{cases}
(1+x)^\gamma & \gamma > 0\\
-(1+x)^\gamma & \gamma < 0.
\end{cases}
\end{align}
Here $f_\gamma(x) \simeq 1 + \gamma x$ as $x \to 0$ (the scaling factor $\gamma$ and constant term is irrelevant for our discussion). As $\gamma$ approaches zero from above, $s[f_\gamma]$ becomes larger, and is infinite in the limit $\gamma\to0$. 
Since $s[f_\gamma]$ implies a rescaling of $a$, as can be seen in \eq{eq:intro:dx2_sf}, the exceedingly large values of $s[f_\gamma]$ for small $\gamma$ means we need a correspondingly large number of lattice sites to approach the continuum limit. 
In any case, as long as $s[f_\gamma]$ implies a finite rescaling, we expect the fractal dimension to be invariant.
For $\gamma <0$, however, the integral $s[f_\gamma]$ does not exist as $a$ approaches zero.

\begin{figure}
\includegraphics[width=\columnwidth]{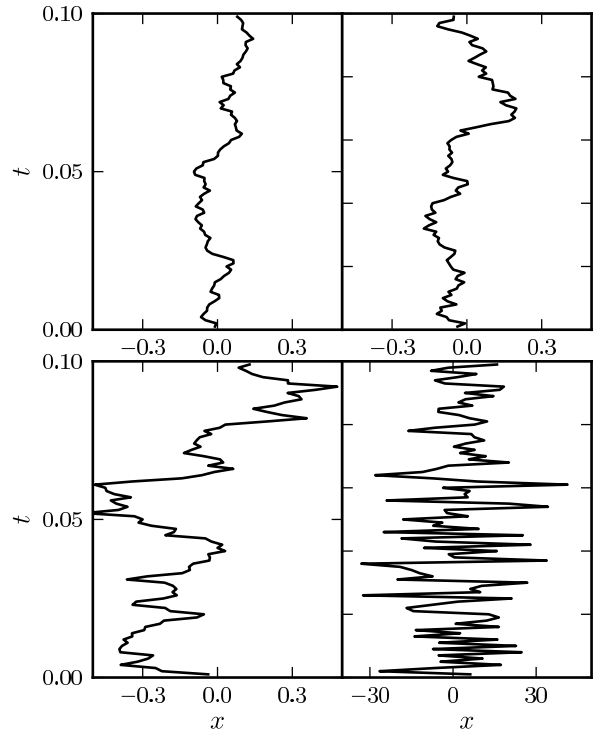}
\caption{\label{fig:intro:falpha} Example paths for the lattice modification defined through \eq{eq:intro:falpha}, for different $\gamma$, showing the various behaviour. Top left is for $\gamma=2.0$, top right for $\gamma = 1.0$, bottom left for $\gamma = 0.5$ and bottom right for $\gamma = -1$, using dimensionless units.}
\end{figure}

\begin{figure}
\includegraphics[width=\columnwidth]{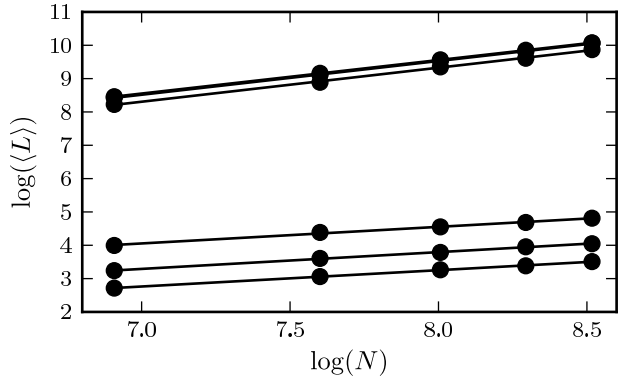}
\caption{\label{fig:intro:LvN}Scaling of the average length $\braket{L}$ as a function of lattice sites $N$. $\log(\braket{L})$ was fitted to $\beta \log(N) + b$. The $\gamma=-1$ and ``tanh'' action coincide at the top line, the second line is for the ``sin'' action, the third for $\gamma=0.5$, the forth for $\gamma=1.0$ and the fifth for $\gamma=2.0$.}
\end{figure}

In \fig{fig:intro:falpha} we show  example paths for a free particle, $V(x) = 0$,  and four different $\gamma$, generated by standard Monte Carlo methods. The paths exhibit larger jumps for smaller $\gamma$. For $\gamma=-1$ the path has a radically different geometry. In \fig{fig:intro:LvN} the length $\braket{L}$ is plotted for varying number of lattice sites, $N$, for the same values of $\gamma$. The scaling $\braket{L} \propto N^\beta$ was found to be $\beta = 0.499\pm0.001$, $\beta = 0.495\pm0.001$, $\beta = 0.495\pm0.004$ and $\beta = 0.997\pm0.003$ for the respective cases $\gamma=2$, $\gamma=1$, $\gamma=0.5$ and $\gamma=-1$ (the errors are mean square errors from the linear regression). The corresponding fractal dimensions, as defined in \eq{eq:intro:df_def}, are consistent with $d_f=2$ for the $\gamma>0$ cases and $d_f=\infty$ for $\gamma=-1$.

The behavior for negative $\gamma$ is in fact typical for \emph{any} modification of the form \eq{eq:intro:fS} with a bounded $f(x)$, and $f(x) = x$ as $x\to0$. In \fig{fig:intro:LvN} we also include results for the modifications $f(x) = \tanh(x)$ and $f(x) = \sin(x)$. The scaling was found to be $\beta = 1.006\pm0.011$ and $\beta = 0.982 \pm 0.007$ respectively, corresponding to an infinite fractal dimension in both cases.

\begin{figure}
\includegraphics[width=\columnwidth]{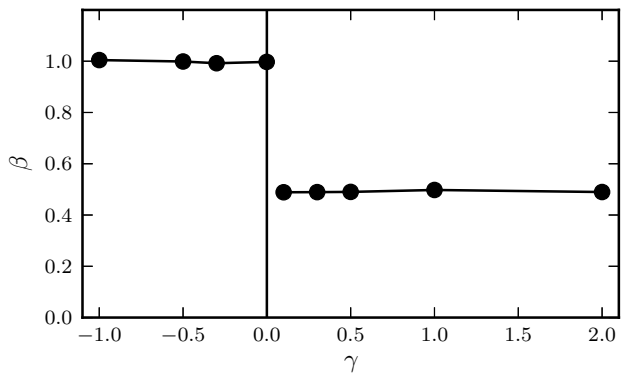}
\caption{\label{fig:intro:fit} $\beta = (d_f-1)/d_f$ as a function of $\gamma$ for the modification \eq{eq:intro:falpha}. For negative $\gamma$, $\beta = 1.0$ corresponds to $d_f = \infty$, and for positive $\gamma$, $\beta = 0.5$ to $d_f = 2$.}
\end{figure}

In \fig{fig:intro:fit} we show how $\beta$ scales with $\gamma$ for the modifications in \eq{eq:intro:falpha}. As $\gamma$ becomes small and positive, there are numerical difficulties due to the necessity of a large number of lattice sites. We here show results for positive $\gamma$ no smaller than $\gamma=0.3$. The results are consistent with $\beta=1$ and $d_f = \infty$ for $\gamma<0$ and $\beta=0.5$, and $d_f=2$ for $\gamma>0$, and points towards critical behaviour at $\gamma=0$, in the limit $N\to\infty$. 

We have also calculated the jaggedness for sub- and super-diffusive actions. In the sub-diffusive case, i.e. actions of the form given in \eq{eq:intro:Sv}, we find results consistent with $J=1/2$ as expected, since there are no correlations between increments $\Delta x_k$ and $\Delta x_{k+1}$ introduced through the modification. This highlights the fact that a classification in terms of jaggedness is independent of a classification in terms of fractal dimension, as was stressed in Ref.\cite{Bogojevic2}. Indeed, even when the paths have a fractal dimension close to one, they are not at all smooth and still fall in to the same jaggedness class, with $J=1/2$.

\begin{figure}
\includegraphics[width=\columnwidth]{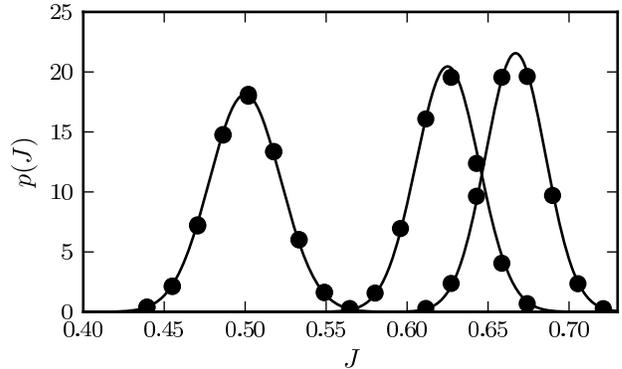}
\caption{\label{fig:intro:J}Typical distributions for the jaggedness with $N=512$ and $a=1/N$. The leftmost Gaussian is centered at 0.5 with width 0.022, and well approximates the case of the naïve action \eq{eq:intro:naive}, and the action given in \eq{eq:intro:Sv} with $\xi=1$ and $\alpha=10$, for which the numerical results represented as dots are nearly indistinguishable. The middle Gaussian is centered at 0.625, and the coinciding dots are numerical results for the action given through \eq{eq:intro:falpha}, with $\gamma=-1$ and the particle's position restrained to a box of width one. The rightmost Gaussian is centered at 0.667, and the dots are numerical results for a uniform distribution of the particles position at each time step.}
\end{figure}

For the super-diffusive case  there is, however,  a subtlety involved in that the particle will always be subject to the infrared boundary effects. In practice, for a finite number of Monte Carlo samples stored on a computer, the particle's position is always confined to some interval for all times, say $-L/2 < x_k < L/2$. If the probability density for the particle's position at time $t_k \simeq ka$, $p(x_k) = |\psi(x_k)|^2$, becomes independent of the position at prior times, such as is the case for the super-diffusive paths considered here, the conditional probability $p_\text{peak}(x_k)$ for a ``peak'' at $x_k$, where a ``peak'' is defined as a point $x_k$ such that $\Delta x_{k-1}$ and $\Delta x_k$ have opposite signs, is just
\begin{align}
&p_\text{peak}(x_k) \equiv 
P{\big(} (x_{k-1}<x_k\, \text{and}\, x_{k+1}<x_k) \nonumber \\
& \text{or}\, (x_{k-1}>x_k\, \text{and}\, x_{k+1}>x_k){\big)} \, .
\end{align}
that is,
\begin{align}
&p_\text{peak}(x_k) = P(x_{k-1}<x_k) P(x_{k+1}<x_k)\nonumber \\
& ~~~~~~~~~~+ P(x_{k-1}>x_k) P(x_{k+1}>x_k)\, .
\end{align}
Consider now, as an example, the case of a uniform distribution on the interval, \ie $p(x_k) = 1/L$ for $-L/2 < x_k < L/2$, and zero otherwise. One easily finds that $p_\text{peak}(x_k) = (1/2 + x_k/L)^2 + (1/2 - x_k/L)^2$. One might think that for large $L$ the probability for a peak should be close to 1/2, but since there is no restriction on the particle's position, it can be close to the boundary for \emph{any} $L$. The expected number of peaks then becomes $\int_{-L/2}^{L/2} p_\text{peak}(x_k)p(x_k)\dx{x_k} = 2/3 \simeq 0.667$, which is, of course, precisely the jaggedness. For super-diffusive actions we find, in numerical simulations, that the jaggedness takes values in between the value for the naïve action and the value for a uniform distribution as just discussed.
In \fig{fig:intro:J} we show some typical example distributions $p(J)$ of the jaggedness. We compare the sub-diffusive case with the usual naïve action, and find that the distributions are nearly indistinguishable and very well approximated by a Gaussian centered at $0.5$. We also show a distribution for a super-diffusive action, and for a uniform distribution of the particle's position at each time step, for comparison.

\section{Conclusions}

To conclude, we have shown that lattice actions, that approach the classical action in the naïve continuum limit, can display highly anomalous behaviour when quantum fluctuations are taken into account. Not only can non-vanishing local potentials be induced by such lattice modifications, as was shown by Klauder et al. in Ref.\cite{Klauder}, but non-local effects can appear in that the geometry of the paths is changed. We have demonstrated modified lattice theories where the paths in the path-integral, with measure greater than zero, exhibit both sub-diffusive and super-diffusive behaviour. We find it  noticeable  that under certain assumptions, a large class of modified actions can, through a renormalization procedure, always be written formally on the naïve discretized form. Finally,  we observe that alternative views on the notion of fractal dimensions in quantum physics has been discussed in the literature as in, \emph{e.g.}, Ref.\cite{Laskin_2000} which, however, is closely related to the notion of fractional derivatives \cite{Metzler_2000} and therefore different from the local deformations of the lattice actions as consider in the present paper.

\appendix*
\section{Renormalizations Induced by Modified Lattice Actions\label{app:renorm}}
For the convenience of the readers we give  a derivation of the Schrödinger equation for the modified mechanics defined in \eq{eq:intro:fS}. Consider the evolution of the wave function over a small time step $a$:
\begin{align}
\psi_{k+1}(x_{k+1}) = N \int_{-L+x_{k+1}}^{L+x_{k+1}} \dx{x_k} \exp(-f(S_{k+1,k})) \psi_k(x_k),
\end{align}
where $N$ is a normalization constant and we have restricted the particles movement to the interval $-L < x_{i+1}-x_i < L$ to ensure the integral always is finite. Introducing the variables $x$ and $y$ through $x = x_{k+1}$ and $x_k = x + \sqrt{a} y$, and by Taylor expanding a sufficiently smooth potential $V(x_k)$, $f(x_k)$ and $\psi_k(x_k)$, dropping terms of order $O(a^2)$, we obtain
\begin{align}
& \psi_{k+1}(x) = N \sqrt{a}  \int_{\Omega} \dx{y} \exp\left\{ -f\left( \frac{y^2}{2} \right) \right\} \psi_k(x) \nonumber \\
& + \frac{1}{2} N \sqrt{a} a \int_{\Omega} \dx{y} y^2 \exp\left\{ -f\left( \frac{y^2}{2} \right) \right\} \psi''_k(x) \\
& - N \sqrt{a} a V(x) \int_{\Omega} \dx{y} f'\left(\frac{y^2}{2}\right) \exp\left\{ -f\left( \frac{y^2}{2} \right) \right\} \psi_k(x) \nonumber ,
\end{align}
 with a domain of  integration $\Omega$ as given by $-L/\sqrt{a} < y < L/\sqrt{a}$. We now choose the normalization constant such that
\begin{align}
N \sqrt{a}  \int_{\Omega} \dx{y} \exp\left\{ -f\left( \frac{y^2}{2} \right) \right\} = 1\, .
\end{align}
Then
\begin{align}
\psi_{k+1}(x) = \psi_k(x) + \frac{a}{2}s[f]\psi''_k(x) - a V(x)g[f] \psi_k(x)\, ,
\end{align}
where $s[f]$ and $g[f]$ are given in Eqs.(\ref{eq:intro:sfgf1}) and (\ref{eq:intro:sfgf2}). We now obtain the following imaginary-time Schrödinger equation
\begin{align}
\lim_{a \to 0} \frac{\psi_{k+1}(x) - \psi_k(x)}{a} =& \pder{\psi(x,t)}{t} \\
=& \frac{1}{2} s[f] \psi''(x,t) - g[f]V(x)\psi(x,t) \nonumber ,
\end{align}
\emph{i.e.},
\begin{align}
\pder{\psi(x,t)}{t} =& \frac{s[f]}{2} \psi''(x,t) - g[f]V(x)\psi(x,t)\, .
\end{align}
Introducing a mass $m$ and $\hbar$ again, we see that $s[f]$ and $g[f]$ constitutes a renormalization of the mass and potential respectively. One can also use this wave equation to show that the imaginary-time commutation relation $[x,p] = \hbar$  still holds in the discretized theory when we use $m_R \dot{x}$ for the momentum and the bare mass $m$ has been replaced by the renormalized mass $m_R = m/s[f]$ (see Section 7-5 in Ref.\cite{FeynmanHibbs}). Since $\hbar$ is unrenormalized we can make use of units such that $\hbar = 1$.


\vspace{0.5cm}
\begin{center}ACKNOWLEDGMENTS
\end{center}


     This work has been supported in part by the Norwegian University of Science and Technology (NTNU) and in part for B.-S.S. by the Norwegian Research Council  under Contract No. NFR 191564/V30, ''Complex Systems and
SoftMaterial'' and the National Science Foundations under Grant. NSF PHY11-25915. The authors are grateful for the hospitality shown at the University of Auckland (ALG), the Center of Advanced Study - CAS - Oslo, KITP at the University of Santa Barbara and B. E. A. Saleh at CREOL, UCF (B.-S.S.), when the present paper was in progress. J.R.K. and B.-S.S. are also grateful to the participants of the 2009 joint NITheP and Stias, Stellenbosch (S.A.), workshop for discussions. A private communication with C. B. Lang on the subject is also appreciated.


\end{document}